\documentclass[prb,aps,superscriptaddress,twocolumn,showkeys,floatfix,citeautoscript]{revtex4-1}

\usepackage{times}
\usepackage{graphicx}

\begin{document}
\title{Real Space Imaging of the Verwey Transition at the (100) Surface of Magnetite}

\author{Juan de la Figuera}
\affiliation{Instituto de Qu\'{\i}mica-F\'{\i}sica ``Rocasolano'', CSIC, Madrid 28006, Spain}
\email{juan.delafiguera@iqfr.csic.es}
\author{Zbynek Novotny}
\affiliation{Institute of Applied Physics, Vienna University of Technology, Vienna, Austria}
\author{Martin Setvin}
\affiliation{Institute of Applied Physics, Vienna University of Technology, Vienna, Austria}
\author{Zhiqiang Mao}
\affiliation{Tulane University, New Orleans, Louisiana 70118, USA}
\author{Gong Chen}
\affiliation{Lawrence Berkeley National Laboratory, Berkeley, California 94720, USA}
\author{Alpha T. N'Diaye}
\affiliation{Lawrence Berkeley National Laboratory, Berkeley, California 94720, USA}
\author{Michael Schmid}
\affiliation{Institute of Applied Physics, Vienna University of Technology, Vienna, Austria}
\author{Ulrike Diebold}
\affiliation{Institute of Applied Physics, Vienna University of Technology, Vienna, Austria}
\author{Andreas K. Schmid}
\affiliation{Lawrence Berkeley National Laboratory, Berkeley, California 94720, USA}
\author{Gareth S. Parkinson}
\affiliation{Institute of Applied Physics, Vienna University of Technology, Vienna, Austria}

\begin{abstract}

Effects of the Verwey transition on the (100) surface of magnetite were studied using scanning tunelling microscopy and spin polarized low-energy electron microsccopy. On cooling through the transition temperature $T_\mathrm V$, the initially flat surface undergoes a roof-like distortion with a periodicity of $\sim$0.5 $\mu$m due to ferroelastic twinning within monoclinic domains of the low-temperature monoclinic structure. The monoclinic c axis orients in the surface plane, along the $[001]_c$ directions. At the atomic scale, the charge-ordered ($\sqrt{2}\times\sqrt{2}$)R45$^\circ$ reconstruction of the (100) surface is unperturbed by the bulk transition, and is continuous over the twin boundaries. Time resolved low-energy electron microscopy movies reveal the structural transition to be first-order at the surface, indicating that the bulk transition is not an extension of the Verwey-like ($\sqrt{2}\times\sqrt{2}$)R45$^\circ$ reconstruction. Although conceptually similar, the charge-ordered phases of the (100) surface and sub-$T_\mathrm V$ bulk of magnetite are unrelated phenomena.
\end{abstract}

\maketitle

Magnetite\cite{CornellBook}, the oldest known magnetic material, has played a central role in the development of modern material science. In the room temperature (RT) cubic phase, magnetite is a half-metallic ferrimagnet\cite{friak_ab_2007} ($T_C$ = 858 K) and a candidate material for spintronics applications\cite{bibes_oxide_2007,wada_efficient_2010}. Cooling through $\sim$120~K (Verwey temperature, $T_\mathrm V$) a first-order phase transition known as the Verwey transition\cite{Verwey,walz_verwey_2002,garcia_verwey_2004} occurs and the system enters an insulating, ferroelectric\cite{kato_observation_1982,miyamoto_measurement_1993,ZieseJPc2012} and ferroelastic\cite{KasamaEarthPlanSciLett2010,SaljeAnnRevMat2012,KasamaPhaseTrans2013} phase with monoclinic $Cc$ symmetry\cite{WrightPRB2002}. The mechanism underlying the Verwey transition remains controversial; there are conflicting reports as to whether the electronic transition is driven by the lattice distortion\cite{garcia_verwey_2004,RozenbergPRL2006}, or vice versa \cite{walz_verwey_2002,WengPRL2012}. In addition to the fundamental interest in phase transitions, novel device concepts utilizing the metal-insulator transitions have been proposed\cite{inoue_taming_2008,wu_magnetization_2013}. The emergence of the monoclinic phase at the Verwey transition was recently observed in the bulk by transmission electron microscopy (TEM \cite{KasamaEarthPlanSciLett2010,KasamaPhaseTrans2013}). 24 equivalent domains are possible since the monoclinic $[001]_m$ direction orients along one of the cubic $[001]_c$ directions. Within each domain, ferroelastic twinning occurs. To date, no real-space images of the magnetite (100) surface below the Verwey transition have been reported. 

Theoretical calculations predict that the magnetite (100) surface remains in the sub-Verwey phase above RT\cite{LodzianaPRL2007}.
This idea is consistent with quantitative low-energy electron diffraction measurements\cite{PentchevaSS2008} which found little change in I-V curves acquired at 78, 200 and 300 K, and scanning tunneling microscopy\cite{WiesendangerScience1992,SchvetsPRB2002,StoltzUltra2008,ParkinsonPRB2010}/spectroscopy\cite{ShvetsPRB2006} measurements, which observe a lattice distortion and a 0.2 eV band gap at room temperature,  respectively. However, low-energy ion scattering studies\cite{BoermaJMMM2000,BoermaSS2001} report changes in the backscattered yield at $\sim$120~K, and angle resolved photoemission experiments detected the opening of a 70~meV bandgap\cite{RanJPd2012}. In this paper we confirm that the Fe$_3$O$_4$(100) surface does not undergo the Verwey transition at 120 K, and we show that the surface is rumpled by the lattice distortions of the underlying bulk at this temperature.

Our experiments were performed using a synthetic Fe$_3$O$_4$(100) crystal grown by the floating zone method. Resistivity measurements show a sharp Verwey transition at 124~K upon heating and 121~K upon cooling (data not shown), indicative of stoichiometric Fe$_3$O$_4$\cite{walz_verwey_2002}. A flat, clean surface was prepared by short sputtering cycles (typically 10 minutes of Ar$^+$ sputtering at 1.5\,keV) followed by annealing to 920~K in a molecular oxygen background of 2$\times$10$^{-6}$ Torr. The scanning tunneling microscopy (STM) work utilized an Omicron LTSTM instrument in constant current mode. The clean sample was transferred from a connected preparation chamber to the measurement chamber. Images were acquired at RT and 78 K. The base pressure in both chambers was in the 10$^{-11}$ Torr range. Spin polarized low-energy electron microscopy (SPLEEM) \cite{rougemaille_magnetic_2010} measurements were performed in a low-energy electron microscope (LEEM, base pressure in the 10$^{-11}$ Torr range) in bright field mode. The instrument is equipped with a spin-polarized electron source and a spin manipulator to adjust the spin direction of the electron beam with respect to the sample surface. Magnetic contrast is obtained in SPLEEM by calculating the asymmetry between low-energy electron microscopy images acquired illuminating the sample with beams of electrons with opposite spin polarization. In the resulting SPLEEM image, bright (dark) areas indicate that the local surface magnetization has a component parallel (antiparallel) to the spin-polarization direction of the electron beam. As the electron beam spin-polarization can be changed with respect to the sample, the magnetization vector can be determined in real space with nanometer resolution \cite{Ramchal2004PRB}. More details on both the instrument \cite{grzelakowski_new_1994}, the spin-polarization control method \cite{duden_compact_1995} or the vector magnetometric application of SPLEEM \cite{Ramchal2004PRB,FaridPRL2006,FaridNJP2008} can be found in the literature. The temperature is measured by means of a thermocouple attached to a molybdenum plate on which the magnetite sample rests. A gold foil between the plate and the sample ensured good thermal contact. The thermocouple was calibrated by dipping the sample-holder in liquid N$_2$. Sample handling requirements in this instrument result in a less-than ideal thermocouple setup and we understand that the absolute thermocouple voltage is potentially afflicted with a margin of error. However, the relative thermocouple voltage variations we measure during slow temperature changes are expected to reflect relative temperature changes with much better accuracy. A conservative estimate is that the absolute accuracy of our temperature measurements may be of the order of 20~K, while repeatability is better than 4~K. 
\begin{figure}
\centerline{\includegraphics[width=0.5\textwidth]{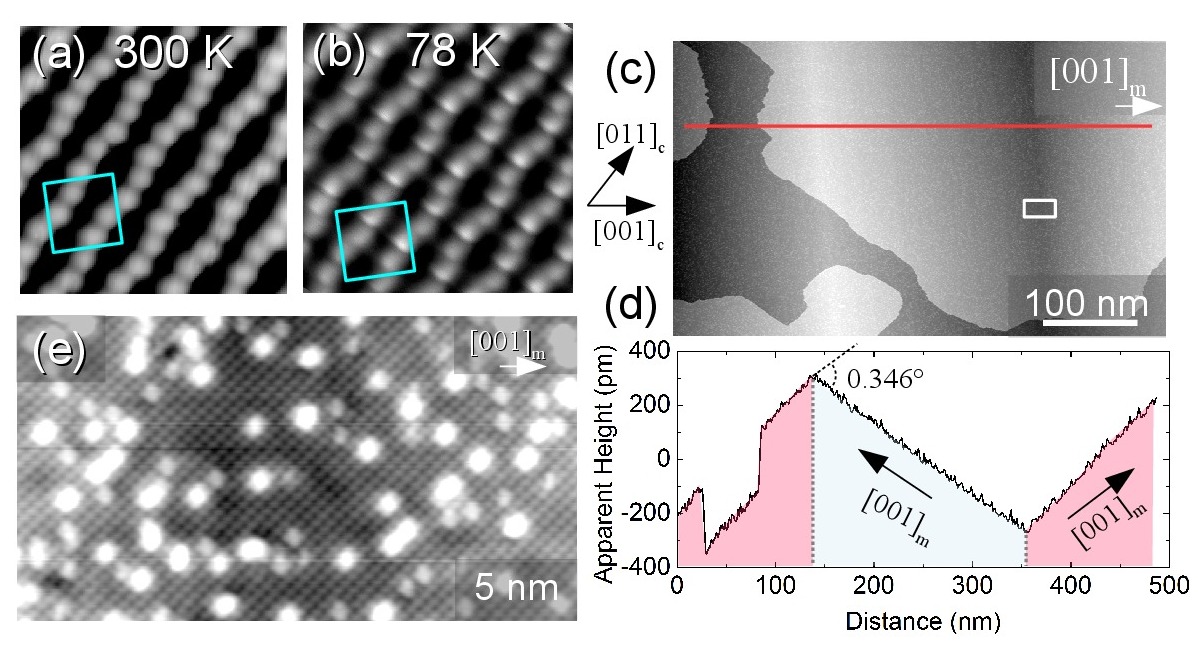}}
\caption{(a) High resolution STM topograph (3.5$\times$3.5~nm$^2$, V$_{sample}=+1V$, I$_{tunnel}=0.3$nA) of the Fe$_3$O$_4$(100) surface acquired at RT, the ($\sqrt{2}\times\sqrt{2}$)R45$^\circ$ unit cell is indicated by the cyan square. (b) High resolution STM topograph (3.5$\times$3.5~nm$^2$, Vsample=+1V, 0.12nA) of the Fe$_3$O$_4$(100) surface acquired at 78 K. The image displays the same ($\sqrt{2}\times\sqrt{2}$)R45$^\circ$ periodicity observed at RT (cyan square).(c) 500$\times$300 nm$^2$ STM image of the Fe$_3$O$_4$(100) surface acquired at 78 K (V$_{sample}=+1$V, I$_{tunnel}=0.1$nA). The surface exhibits a roof like structure with a periodicity of approx. $\sim$0.5 $\mu$m. (d) Profile along the line marked in red in (c). The angle at the ridge is 0.346$^\circ$ (+/-0.018$^\circ$). (e) 30$\times$16.3 nm$^2$ STM image (V$_(sample)=+1$V, I$_{tunnel}=0.1$nA) showing a valley of the roof structure. No disruption of the atomic structure occurs at the twin boundary. }
\label{stm_and_leem}
\end{figure} 

In Fig. \ref{stm_and_leem}(a,b) we show atomically resolved STM images collected at RT and 78 K. Both images exhibit the undulating Fe rows\cite{WiesendangerScience1992,SchvetsPRB2002,StoltzUltra2008,ParkinsonPRB2010} associated with the ($\sqrt{2}\times\sqrt{2}$)R45$^\circ$ reconstruction, confirming unchanged surface periodicity at the bulk Verwey transition temperature. On a larger scale (500$\times$300 nm$^2$) however, STM topographs acquired at 78~K reveal a roof-like distortion with a periodicity of $\sim$ 500~nm [Fig.~\ref{stm_and_leem}(c)]. Line profiles [red line, Fig.~\ref{stm_and_leem}(c,d)] show an angle between the planes of 0.346$^\circ$ (+/-0.018$^\circ$). The ridges and valleys run along the $[001]_c$ crystal direction and no correlations with morphological features such as surface steps are apparent. High-resolution STM topographs acquired across the ridges and valleys of the roof exhibit no evidence of any discontinuity in the surface structure at the ridges or valleys [the area imaged in (e) shows a valley and corresponds to the white rectangle indicated in (c)].

\begin{figure}
\centerline{\includegraphics[width=0.4\textwidth]{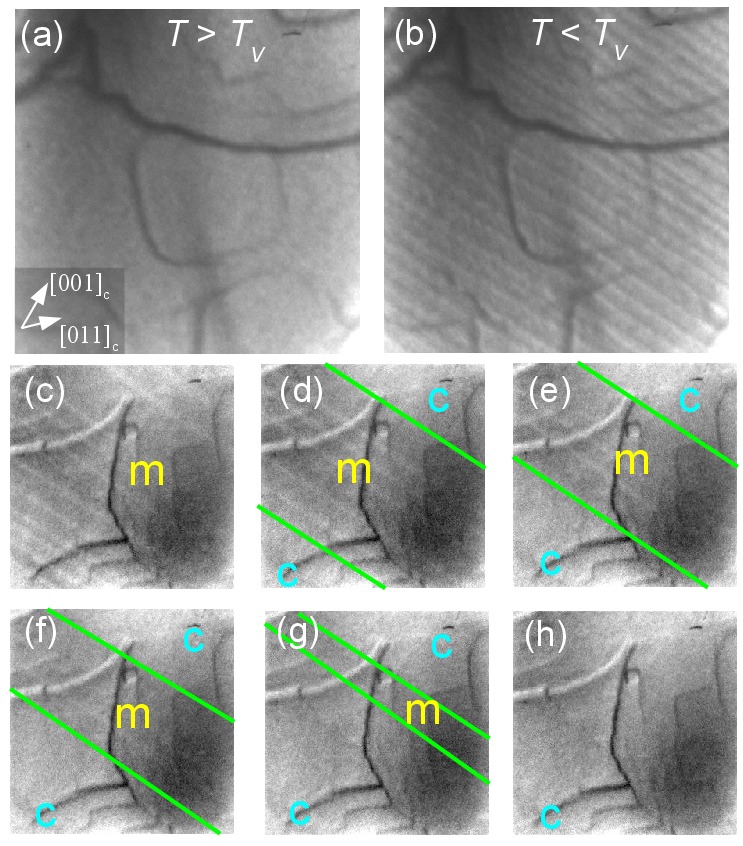}}
\caption{Imaging the Verwey transition at the Fe$_3$O$_4$(100) surface using LEEM. The size of all the images is 8.6$\times$8.6 $\mu$m$^{2}$, and the electron energy is 9.4 eV. (a) LEEM image of the Fe$_3$O$_4$(100) surface acquired at room temperature. Dark lines in the image correspond to step bunches. (b) LEEM image of the sub $T_V$ Fe$_3$O$_4$(100) surface. Light-dark stripes with an average periodicity of $\sim$0.5 $\mu$m run perpendicular to the local $[001]_m$ direction of the monoclinic sub-$T_\mathrm V$ structure. (c--h) Selected LEEM images from a LEEM movie (see supplement) acquired while slowly heating the sample from 121 K. Over the course of the movie (duration 20 seconds, temp. increase 0.5 K) the striped phase (m) recedes from the top and bottom of the imaged area and is replaced by the RT (cubic) phase (c). The boundaries between both phases are marked by green lines. }
\label{heating}
\end{figure}

Establishing the role this ridge-valley morphology plays in the Verwey transition requires examining its dynamic temperature dependence, as well as confirming its structure across larger regions of the surface. To this end, we complement the high spatial resolution STM results with the larger field of view and better time-resolution of variable temperature LEEM measurements. 
In Fig.~\ref{heating} we show LEEM images acquired well above (a), well below (b), and while slowly heating through the Verwey transition, (c--h).  
At RT, the LEEM image (a) exhibits dark lines associated with surface step bunches. In addition, on cooling through the Verwey transition, rows of parallel alternating bright and dark stripes appear, (b). No correlation between the stripes and step bunches is observed, rather the orientation of the stripes is dictated by the crystallographic directions. The stripes are almost always oriented along an in-plane $[001]_c$ direction of the cubic unit cell, and very rarely we find regions where they are oriented along $[011]_c$. The stripes are typically a fraction of a micrometer in width, disappear upon heating above $T_\mathrm V$ and reappear again when cooling. The fast acquisition time of the LEEM allows for real-time imaging through the transition. Fig.~\ref{heating} (c--h) shows selected frames from a longer LEEM movie (see supplement, 0.5 s per frame) acquired while slowly heating the sample through the transition temperature. The striped phase recedes from the top and bottom of the imaged area, finally disappearing in the center. The time and temperature span of the movie is 20 seconds and 0.5 K, respectively. Again, the surface topography has no effect on the propagation on the striped phase. The thermocouple reading indicates that the Verwey transition happens at 105$\pm4$~K when cooling down and at 121$\pm2$~K when heating up. Given the thermocouple reading limitations in the SPLEEM setup as mentioned above, this is in reasonable agreement with the resistivity measurements on the same sample.

\begin{figure}
\centerline{\includegraphics[width=0.4\textwidth]{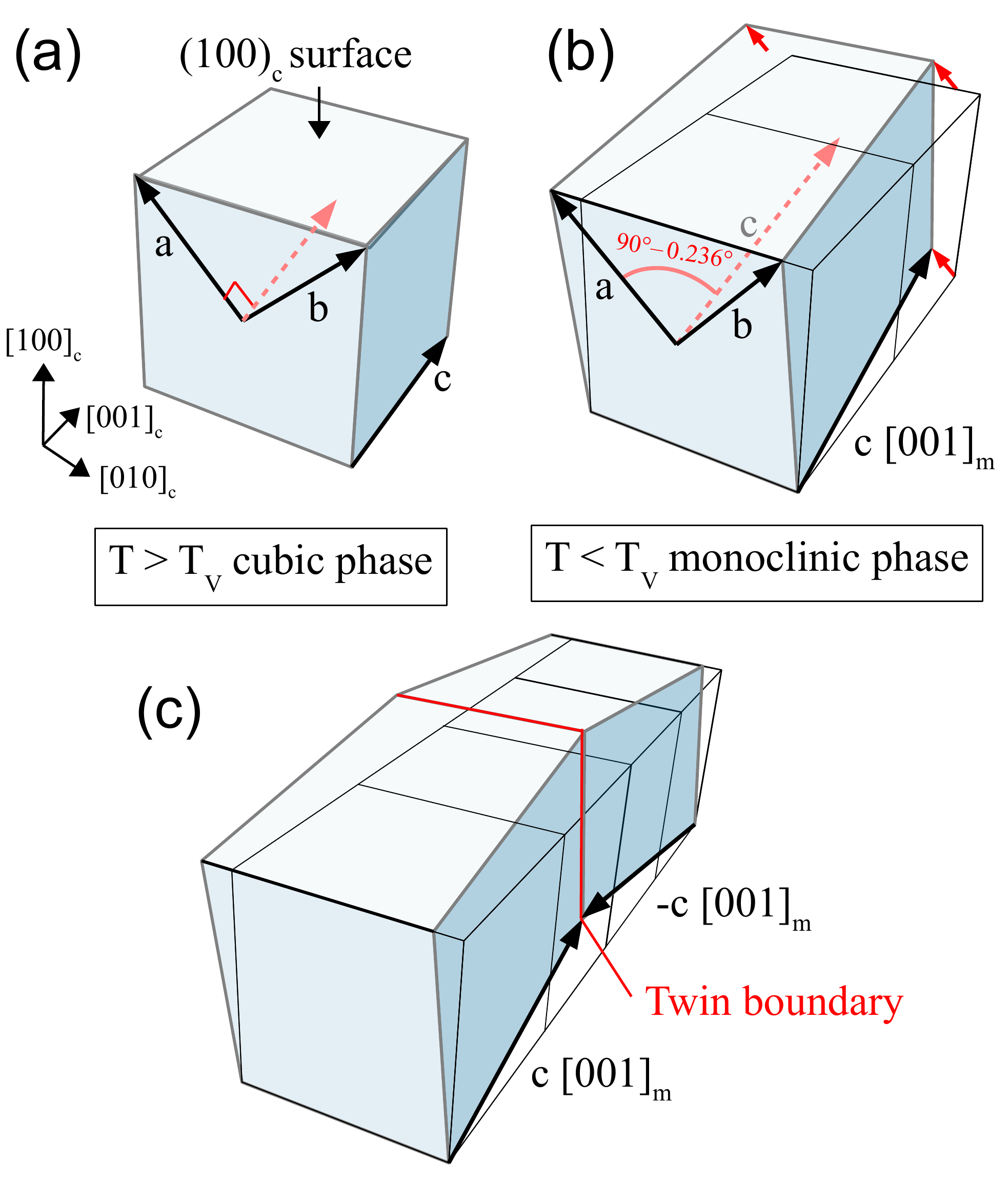}}
\caption{Schematic illustrating the lattice distortions that occur in the Verwey transition in magnetite. (a) The cubic unit cell of magnetite in the high temperature phase drawn in perspective view. (b) The low-temperature monoclinic unit cell of magnetite. A lattice distortion causes the lattice vectors {\bf a} and {\bf b} to become inequivalent. The periodicity in the {\bf c} direction doubles to the equivalent of two cubic unit cells. The angle between the {\bf a} and {\bf c} vectors distorts from 90$^\circ$ to 89.764$^\circ$, causing the top (100) surface to tilt with respect to the high temperature cubic cell. (c) Two mirrored monoclinic cells with common but opposite monoclinic {\bf c} axis joined at a twin boundary (indicated by the red lines). The top (100) surfaces meet at a peak at the twin boundary.}
\label{schematic}
\end{figure}

The roof-like ridge-valley structure observed at the Fe$_3$O$_4$(100) surface can be understood by considering the distortions that occur during the Verwey transition. In Fig.~\ref{schematic}(a), the cubic unit cell of the RT phase is sketched in perspective view. The surfaces of the cube are (100) type surfaces. In the low temperature phase, see Fig.~\ref{schematic}(b), the {\bf a} and {\bf b} vectors become 5.94441 \AA{}ngstrom and 5.92472 \AA{}ngstrom respectively\cite{IizumiSSCOM1975,GasparovPRB2000,MiyamotoPRB1999,WrightPRB2002} and the periodicity doubles in the monoclinic c direction (monoclinic $[001]_m$) to 16.77508 Angstrom. The angle between the {\bf a} and {\bf c} vectors distorts to 89.764$^\circ$, inclining the (100) surfaces parallel to the [001]$_m$ axis with respect to the RT cubic plane by approximately 0.17$^\circ$. Fig.~\ref{schematic}(c) shows how two monoclinic unit cells that share an aligned, but opposite monoclinic {\bf c} axis, can be joined forming a twin. The twin boundary (red lines) is a local mirror plane, with minimal disruption of the structure across the boundary. On the upper (100) surface, the distortion would appear as a ridge with an angle of 0.32$^\circ$, close to that measured in our STM images. Therefore we conclude that the ridge-valley structure observed in the STM and LEEM data is related to the Verwey transition in magnetite, and that the [001]$_m$ axis in the vicinity of the Fe$_3$O$_4$(100) surface is oriented in-plane, and aligned with one of the $[001]_c$ axes of the RT phase. 

\begin{figure}[h!]
\centerline{\includegraphics[width=0.45\textwidth]{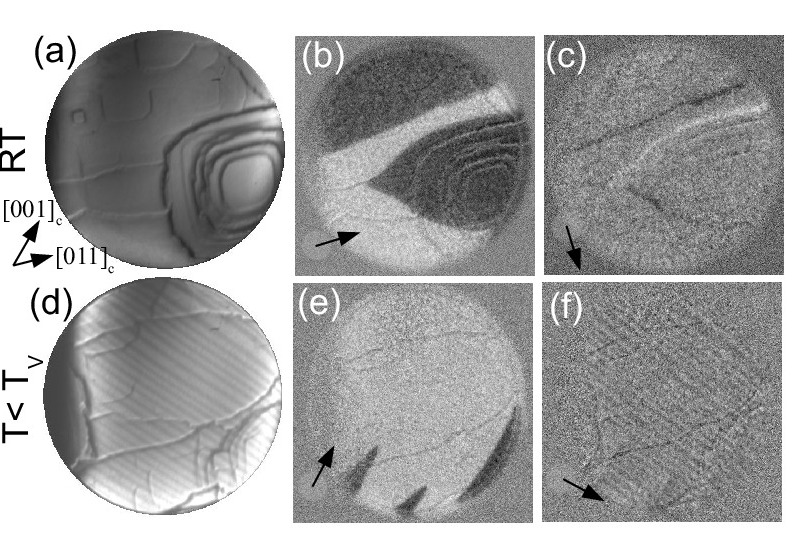}}
\caption{(a) LEEM image of the Fe$_3$O$_4$(100) surface at RT. (b--c) SPLEEM images showing the magnetization along $[011]_c$ and $[01\bar1]_c$  respectively. (d) LEEM image of the Fe$_3$O$_4$(100) surface below Tv. Light/dark stripes due to differing electron reflectivity from microtwins run parallel to the the $[001]_c$ directions. (e--f) SPLEEM images showing the magnetization along $[001]_c$ and $[010]_c$ directions. No out of plane magnetization is observed, neither at RT or below Tv. All the images have a field of view of 13~$\mu$m, and are acquired at a start voltage of 9.3~eV.}
\label{mag}
\end{figure}

We can test this model by looking at its implications regarding magnetic properties. The RT easy axes of magnetization in Fe$_3$O$_4$(100) bulk material are the $\langle111\rangle_c$  type directions\cite{MuxworthyGeoPhysJInt2000}. At the (100) surface, the dipolar interaction exceeds the relatively weak magnetocrystalline anisotropy of the RT phase and as a result the closest in-plane directions, $[011]_c$ and $[0\bar 11]_c$, have been found to be the easy magnetization axes\cite{AndreasUltra2013}. Below $T_\mathrm V$, prior work shows that the easy magnetization axis is the $[001]_m$ direction \cite{MuxworthyGeoPhysJInt2000}. SPLEEM images of magnetic domain patterns in the magnetite surface are consistent with this prediction. The magnetic domains on the surface at RT and below $T_\mathrm V$  are shown in Fig. 4, where we present magnetic-contrast images of the surface along  orthogonal  directions  within the surface plane.   In the top row RT topography and the magnetic-contrast images along in-plane $[011]_c$ directions are shown (no significant contrast was detected in the out-of-plane direction, not shown). The spin polarization of the imaging beam was aligned to condition of either maximum magnetic contrast, Fig.~\ref{mag}(b, e), or vanishing magnetic contrast, Fig.~\ref{mag}(c, f). The images indicate that the easy-axes are along the $[011]_c$ and $[0\bar 11]_c$ directions at RT. Below $T_\mathrm V$ the magnetic contrast vanishes when the electron beam polarization is oriented parallel to the stripes, i. e. along the $[001]_m$ direction as shown in Fig.~\ref{mag}(f). (Weak row-like contrast in this image is due to a small spin canting in the microtwins, of about 8$^\circ$ around the [001]$_m$ direction).

An interesting question is whether the observed roof-like structure is related to a ``surface Verwey transition'', or if the surface is merely distorted by the bulk Verwey transition occurring beneath. Our atomic-resolution STM images acquired above and below the transition show that the characteristic relaxations of the ($\sqrt{2}\times\sqrt{2}$)R45$^\circ$ reconstruction persist across the bulk transition temperature, in agreement with the results of a quantitative LEED study\cite{PentchevaSS2008}. Both results fit with a picture where the ($\sqrt{2}\times\sqrt{2}$)R45$^\circ$ reconstruction is a distorted, charge-ordered state well above $T_V$\cite{LodzianaPRL2007}. Photoemission experiments detected the opening of the bandgap at $T_V$\cite{RanJPd2012}, but this could be explained by the fact that this technique is not inherently surface sensitive, but instead probes a mixture of surface and bulk properties. Extremely surface sensitive STS measurements observed a similarly sized band gap already at RT\cite{ShvetsPRB2006}. Consequently, we contend that the observed Verwey-roof at the Fe$_3$O$_4$(100) surface is a consequence of the bulk Verwey transition, and is not related to a transition in the surface layers.

One might expect that if the ($\sqrt{2}\times\sqrt{2}$)R45$^\circ$ reconstruction at the surface was indeed a ``surface Verwey transition'', as predicted by recent theoretical calculations\cite{LodzianaPRL2007}, it would act as a nucleation site for the bulk Verwey phase. In such case, the surface would be expected to be the last portion of the crystal with the monoclinic phase to disappear. Taking the roof-like distortion as an order parameter, we clearly see a first-order phase transition at the surface, in agreement to what is observed for the bulk Verwey transition. This cannot be reconciled with a gradual extension of order inward from the surface. In any case, the ($\sqrt{2}\times\sqrt{2}$)R45$^\circ$ reconstruction differs significantly from the monoclinic bulk unit cell, both in terms of its overall symmetry (smaller unit cell) and the proposed charge-order of the subsurface octahedral Fe rows (Fe$^{2+}$-Fe$^{2+}$-Fe$^{3+}$-Fe$^{3+}$). Therefore, the surface and bulk ordering are unrelated, but both arise from the propensity of magnetite to form distorted, orbital-ordered phases. It would be interesting to investigate whether the ($\sqrt{2}\times\sqrt{2}$)R45$^\circ$ reconstruction (or possibly another charge ordered state) is retained at the interface between Fe$_3$O$_4$(100) and other oxide materials. If so, the presence of an insulating ``dead layer'' could explain the unexpectedly poor performance of Fe$_3$O$_4$ based spintronics devices\cite{MoussyJPhysd2013}.

In summary, we have combined surface (STM and LEEM) and magnetic imaging techniques (SPLEEM) to study magnetic and structural effects of the bulk Verwey transition at the (100) surface of magnetite. We find that the transition to the low-temperature phase is associated with the formation of a ridge-valley topography at the surface. This topography is interpreted to result from the microtwin structure within the domains of the monoclinic, ferroelastic phase of magnetite. These results show that in the vicinity of the Fe3O4(100) surface the monoclinic {\bf c} axis preferentially orients in-plane along $[001]_m$ type directions and, since the monoclinic {\bf c} axis is also the magnetic easy axis, magnetization in the low temperature phase is also along these directions. High-resolution STM images show no change in the surface structure across $T_\mathrm V$, and no disruption of the surface structure is observed at the twin boundaries. Both results are consistent with a model where the ($\sqrt{2}\times\sqrt{2}$)R45$^\circ$ surface remains in an independent Verwey-like state well above the bulk transition temperature.

This research was supported by the Spanish Government through projects Nos.~MAT2009-14578-C03-01 and MAT2012-38045-C04-01, by the Office of Basic Energy Sciences, Division of Materials and Engineering Sciences, U.~S. Department of Energy under Contract No. AC02—05CH11231 and by the Centre for Atomic-Level Catalyst Design, an Energy Frontier Research Centre funded by the U.S. Department of Energy, Office of Science, Office of Basic Energy Sciences under Award Number DE-SC0001058. GSP acknowledges support from the Austrian Science Fund project number P24925-N20. 

%

\end{document}